# An Efficient Multi-Link Channel Model for LiFi

Sreelal Maravanchery Mana, Kerolos Gabra Kamel Gabra, Sepideh Mohammadi Kouhini, Peter Hellwig, Jonas Hilt, Volker Jungnickel
*Photonic Networks and Systems, Fraunhofer Heinrich Hertz Institute,*
Einsteinufer 37, 10587 Berlin, Germany
sreelal.maravanchery@hhi.fraunhofer.de

*Abstract*—In this paper, we report for the first time on a new channel modelling technique for multi-link LiFi scenarios. By considering simplified numerical calculation, it models the links between multiple optical frontends and multiple mobile devices much faster than previous approaches. For the first two diffuse reflections, we replace ray tracing method by frequency domain channel modelling technique. For the other higher order diffuse reflections, we use a well-established model based on the integrating sphere. For validation of our new approach, we performed distributed 4x2 MIMO channel measurements. Comparison of simulation and measurement yields a relative mean square error below 5 percent for the signals with a free line-of-sight. Our new technique enables efficient modeling of mobile scenarios and analyzing statistical properties of the LiFi channels.

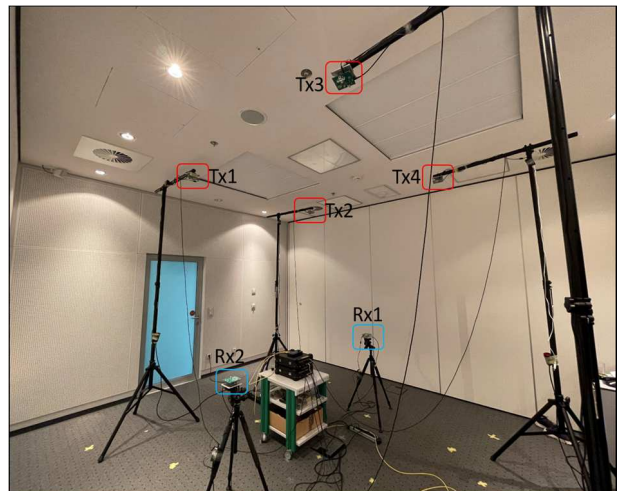

Fig. 1.    Measurement scenario in an empty conference room. Transmitters and receivers are marked in red and blue color.

## I. INTRODUCTION

The emergence of LiFi for indoor communications opens up new possibilities for wireless services in crowded multiuser scenarios [1]. LiFi offers reliable, secure, high data rate wireless links [2]. It is hard to wiretap or jam a LiFi link because signals propagate only inside the room and in small cells defined by the light beam, improving reliability and security. However, the use of LiFi in indoor scenarios may be challenging due to the line-of-sight (LOS) blockage, although connectivity is still possible via weak diffuse reflections. A better approach is to deploy multiple transmitters at the ceiling and use another LOS link for resilience in case of blockage. These techniques are also referred to as multiple-input multiple-output (MIMO).

Channel modelling is an important step when developing any communication system, and it is needed for performance evaluation [3]. A comprehensive understanding of the channel and an accurate prediction of the link are indispensable to develop an optimal transmission scheme for LiFi.

In most indoor environments, the light is diffusely reflected, and the resulting multi-path signals lead to inter-symbol interference (ISI). The importance of theoretical and experimental analyses of indoor optical wireless channels is evident. Recently, different channel modelling methods were introduced, which can be classified into deterministic [4]-[8] and non-deterministic approaches [9]-[12]. Most of these techniques have not been validated through measurements in real scenarios. Recently, a site-specific characterization based on non-sequential ray tracing has been introduced by using commercial optical design software, Zemax® [13]. This approach is capable of obtaining the channel impulse response for any source while considering both the specular as well as mixed specular-diffuse reflections. However, ray tracing operates in the time domain and it follows each path until it either absorbed or received and modelling MIMO and mobility becomes very complex in this way. In practice, fewer reflections are used because the LOS and first diffuse reflections have the most noticeable effects on the achievable data rate.

For future wireless applications, it is expected that a large number of internet of things (IoT) devices will be connected through LiFi systems by using distributed optical frontends (OFEs). Due to a large number of concurrent links, the existing channel modelling techniques for LiFi become insufficient. A new approach is needed aiming at the so-called system-level analysis with a large number of links including the desired signals and their mutual interference. A feasible channel model should provide the channel parameters of a given indoor environment for a large number of parallel links and support the mobility of the IoT devices as well. Only in this way, it becomes possible to characterize the channel quality for many devices that will communicate in parallel.

For the first time, in this paper, we introduce a simplified channel modelling technique for large-scale LiFi deployments. We combine the frequency domain technique recently introduced by H. Schulze [6] with the well-known integrating sphere model [7]. The combination of both approaches enables efficient planning of future LiFi networks. The main contribution of this paper is to show how both approaches can be combined efficiently and the first validation of this new approach is provided using the measurements in a controlled environment. Measurements have been conducted in a 4x2 MIMO downlink scenario, as shown in Fig.1, with 4 OFEs and 2 mobile devices using a direct current (DC) biased orthogonal frequency division multiplexing (OFDM) based channel-sounding technique. For validation, we compare measurement and simulation results.

The remainder of the paper is organized as follows. Section II describes the efficient LiFi channel modelling methodology and discusses the simulation environment.

Section III introduces the experimental setup, the measurement scenario and configurations of the optical frontends. Section VI reports the simulation results as well as measurement results. A detailed discussion of the results are provided in section V and conclusions are given in section VI.

## II. LiFi Channel Modelling Methodology

The channel modelling simulations are performed in the frequency domain rather than in the time domain. This method calculates channel transfer functions instead of impulse responses. LiFi links are modelled as LOS link and NLOS links. For LOS, we followed the widely accepted model as in [14], which is based on the orientation and optical parameters of Tx and Rx. For NLOS, we calculated responses for the first two diffuse reflections using the frequency domain technique [6] and all higher order reflections using the model based on the integrating sphere [7]. In both LOS as well as NLOS channel modelling scenario, we considered the same optical parameters such as field of view (FOV) of the photodiode (PD) and radiation pattern of the LED same as given in the [13].

### A. LOS Channel Model

The generalized LOS channel model between Tx and Rx is given as

$$H_{LOS}(f) = L_{Tx,Rx} \cdot e^{-j2\pi f \tau_{Tx,Rx}} \quad (1)$$

Here $L_{Tx,Rx}$ is the transfer function coefficient between Tx and Rx which can be described same as in Equation (2) in [15]. This coefficient depends on the FOV of the photodiode and the radiation pattern of LED [14]. The variable $\tau_{Tx,Rx}$ is the delay time which depends on $d_{Tx,Rx}$ being the distance between Tx and Rx and $c$ which is the speed of the light [6].

### B. First two diffuse reflections

The first two diffuse reflections are calculated using the frequency domain method [6]. This method considers the reflecting surfaces as $N$ surface elements. This allows assembling all mutual LOS links between all surface elements and the links between the surface elements to the Rx and Tx in a matrix form so that higher-order reflections can be described by consecutive matrix multiplications [6].

We assume that there are $N$ surface elements in the room. For a single Tx to Rx scenario, as described in [6], the entire diffuse channel model can be represented as

$$H_{diff}(f) = r^T(f) \cdot G_\rho \cdot \sum_{m=0}^{\infty} \left(H(f) \cdot G_\rho\right)^m \cdot t(f) \quad (2)$$

where $m$ is the reflection order, $r^T(f)$ is the LOS transfer functions of the link from each surface element in the room to the receiver, $t(f)$ is the LOS transfer functions for the links from the transmitter to all surface elements, $G_\rho$ is a diagonal $NxN$ reflectivity matrix, where each diagonal element $\rho_k$ represents the reflectivity of the $k^{th}$ surface element. The $NxN$ matrix $H(f)$ is the intrinsic transfer matrix, where each element $H_{i,k}(f)$ in this matrix represents the LOS transfer function from $k^{th}$ surface element to $i^{th}$ surface element. Note that, for simplicity, we consider all $N$ reflecting surface elements as Lambertian surfaces.

In our simulations, Equation (3) is calculated only for the first two reflections using an iterative approach, see section III. D in [6]. For a MIMO scenario, we followed the same approach as given in [6] (see section III. G), where $r^T(f)$ and $t(f)$ is calculated for each Tx and Rx configuration in the room. Here, for a particular indoor scenario, it is required only once to calculate the intrinsic transfer function matrix $H(f)$. In the case of mobile MIMO scenario, furthermore, we needed to calculate $r^T(f)$ corresponding to the movement of users in the room.

### C. Higher order diffuse reflections

From previous channel measurements, we observed that the higher order diffuse reflections have all very similar characteristics [16]. This is intuitive as diffuse reflections illuminate the entire room. This motivates us to consider that higher order diffuse reflections depend more on the environment than on the orientation of Tx and Rx. Higher order diffuse reflections can be modelled altogether by using Ulbricht's integrating sphere model, which has been adapted to regular room dimensions in [7]. In contrast to the microscopic approach described above, this macroscopic model does not include any details of the room except a few basic parameters. The generalized diffuse channel model for a given room model is given by Equation (9) in [7]. Since the first two diffuse reflections are calculated using the previously mentioned frequency-domain method, here we need to calculate only the higher-order diffuse reflections starting from the third order onwards. Thus, the higher order diffuse reflections can be expressed as

$$H_{diff_{high}}(f) = \frac{\eta_{diff-(1,2)}}{1 + j2\pi f \tau} \quad (3)$$

where $\tau$ is the exponential decay time which is related to room parameters as described in [7]. The variable $\eta_{diff-(1,2)}$ is the diffuse channel gain excluding the first two reflections and can be expressed as

$$\eta_{diff-(1,2)} = \rho_1 \frac{A_{Rx}}{A_{room}} \left(\frac{1}{1 - \langle\rho\rangle} - 1 - \langle\rho\rangle\right) \quad (4)$$

and it can be simplified as

$$\eta_{diff-(1,2)} = \rho_1 \frac{A_{Rx}}{A_{room}} \frac{\langle\rho\rangle^2}{1 - \langle\rho\rangle} \quad (5)$$

where $A_{room}$ is the area of the room surface, $A_{Rx}$ is the area of the receiver, $\rho_1$ is the reflectivity of the region initially illuminated by the Tx and $\langle\rho\rangle$ is the average reflectivity of the room, see [7].

This model provides an approximate result for higher order reflections in a given room. However, it lacks precision at the beginning of the impulse response, particularly during the first and second diffuse reflections, which we have calculated for each Tx-Rx link separately. Since higher order diffuse reflections do not dependent on Rx orientations, all calculations have been made only once for a downlink mobile user scenario.

Finally, the complete transfer function of the LiFi channel can be represented as

$$H_{total}(f) = H_{LOS}(f) + H_{diff}(f) + H_{diff_{high}}(f) \quad (6)$$

where $m$ ranges between 0 and 1 in Equation (2). Using Equation (6), we calculate the channel transfer function at each frequency $f$. Note that, in this paper, for simplicity we considered the reflecting surfaces as Lambertian. The model is not limited regarding the surface reflection characteristics.

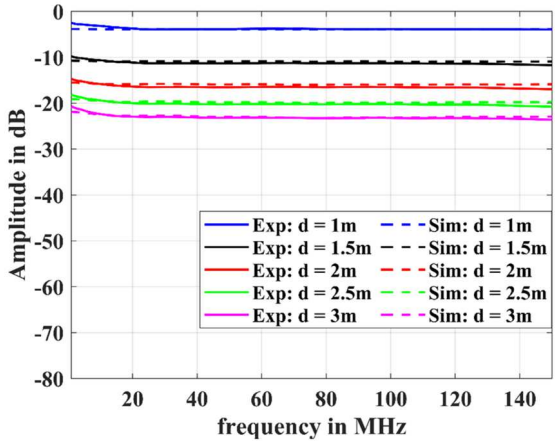

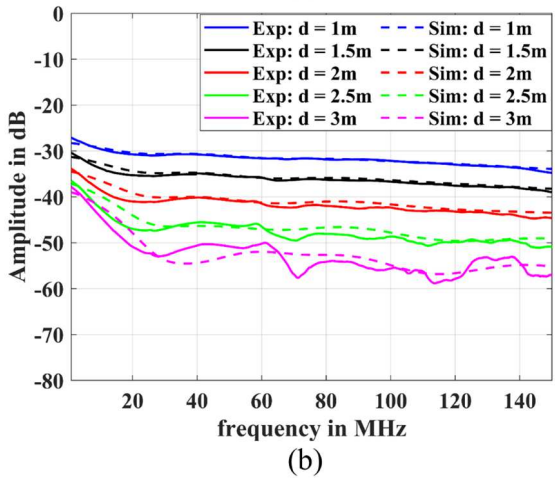

Fig. 2. Amplitude response of the LiFi channels in a SISO scenario. (a) LOS channels, (b) NLOS channels with dominant first order reflections [13].

TABLE I.  MEAN SQUARE ERROR

| Distance of separation between Tx and Rx (in meters) | MSE in percentage | |
|---|---|---|
| | LOS Scenario | NLOS Scenario |
| 1 | 0.1 | 0.1 |
| 1.5 | 0.3 | 0.4 |
| 2 | 0.5 | 1.5 |
| 2.5 | 0.3 | 3.8 |
| 3 | 0.1 | 6.7 |

It is possible to define any surfaces with any reflectance values and characteristics (like Phong reflection model [17]).

## III. LIFI CHANNEL MEASUREMENTS

MIMO channel measurements have been conducted in an empty conference room, with the size of 5.8 m x 4.5 m x 3.1 m as shown in Fig. 1, using a channel sounder system developed in our lab [18]. The LiFi channel sounder is capable of performing broadband 8×8 MIMO channel measurements at frequencies of up to 250 MHz. The widely used multi-carrier approach, DC biased OFDM, is used for simultaneous measurement of MIMO LiFi channels versus frequency as described in [18]. Each frequency response of the LiFi system includes the response of optical frontends, wires and optical propagation channels. All the measurement data are post-processed using the same method reported in [19]. Particularly, frontends and cable responses were calibrated out.

The measurements are conducted in downlink and mobile user scenario. During the measurements, the transmitters are kept in a 2m x 2m grid size, at 2.85 m height. The receivers are kept at 1m height looking towards the ceiling. In the downlink, Rx1 is kept in the center position between Tx2 and Tx4 and Rx2 is in the middle between Tx1 and Tx3. Finally, we considered a mobile user scenario where Rx1 moves around in the room and Rx2 is kept fixed at the nearby center position in the room. Measurements have been done at 40 different positions, where Rx1 is moving along the 2 m x 2 m grid line.

## IV. RESULTS

In this section, we report the measurement and simulation results. By assuming a MATLAB based simplified room model and considering all optical parameters the same as reported in [13], we estimated the channel response in three different configurations such as *i)* SISO, *ii)* 4x2 MIMO and *iii)* a mobility scenario. Note that, in our simulations, we considered only walls, ceiling, and floor of the room where the reflectivity parameters are the same as given in [13]. At first, to validate our new simulation method, we consider the experimental results of the SISO scenarios that we reported in [16]. In the same room, we performed 4x2 MIMO measurements and compared those results with simulation results.

### A. SISO Scenario

The amplitude response of the LiFi channels in a SISO scenario is shown in Fig. 2. As described in [13], here we considered two configurations; LOS and NLOS with dominant first order reflections.

Fig. 2 (a) shows the amplitude response of the LiFi channels at different distances where Tx and Rx are looking each other in a LOS scenario. We observe that the amplitude response is relatively flat overall frequencies. As the separation distance between Tx and Rx increases from 1 m to 3 m, the DC channel gain reduces. From the simulation results, shown as dotted lines, it is clear that the simulated channel response is in good agreement with measured results.

Fig. 2 (b) shows the amplitude response of the NLOS channels where Tx and Rx kept in NLOS configurations at different distances. Both Tx and Rx face towards the ceiling as described in [13]. Here most of the dominant contribution of the signals come from the first order reflection path. As the separation distance increases, the possible first order reflection reduces and higher order reflections get larger [16]. In this case, for the channels with DC channel gain more than -35 dB, simulation results are in good agreement with the experimental results. As distance increases, ripple effects become noticeable at higher frequencies due to the noise, which creates more deviation when compared to the simulation results.

To measure the accuracy of our simulation methodology, we calculate the mean square error (MSE) between measurement and simulation data for both LOS and NLOS channels (shown in Fig. 2). The calculated MSE values are shown in Table I. It is observed that the MSE is less than 2 percentage for LOS scenarios and less than 5 percentage for NLOS configurations with higher channel gain. Note that, the error is increased if the DC channel gain is decreased below -35 dB.

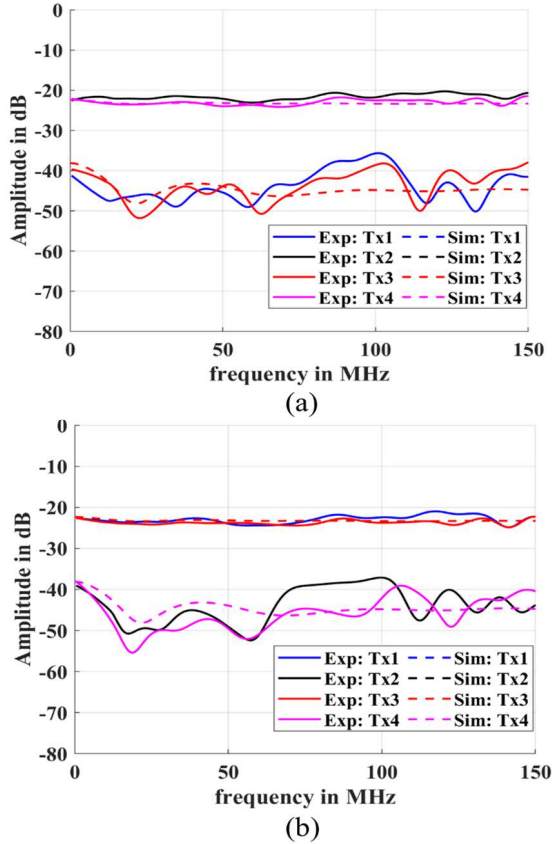

Fig. 3. Amplitude response of the LiFi channels (a) at Rx1 and (b) at Rx2 in a 4x2 MIMO scenario.

*B. 4X2 MIMO Scenario*

The measurement and simulation results of a distributed 4x2 MIMO LiFi (see Fig. 1) are shown in Fig. 3. The channel responses at Rx1, placed in the middle of transmitters Tx2 and Tx4 (see Fig. 1), are shown in Fig. 3 (a). There are strong LOS signals from Tx2 as well as from Tx4 and weak signals from other transmitters Tx1 and Tx3. In Fig. 3, bold lines denote the measured channel responses and dotted lines show the simulated channel responses. Since all transmitters kept in a 2 m x 2 m grid configuration, as shown in Fig. 1, the simulated channel responses between Rx1 to Tx2 and Rx1 to Tx4 will be the same. In the experiment, however, due to small differences in the optical frontends, wires and connectors, the measured responses have minor differences from each other and do not overlap like the simulated curves. We observed that channel responses with respect to the transmitters Tx2 and Tx4 have lower signal strength. So these measurement results could be affected by the noise. That is the reason why we observe that the measured data fluctuate more noticeably when compared to the simulation results. To measure the accuracy of our simulation methodology, we calculate the relative MSE between measurement and simulation data of all the links between each Rx and Tx. The MSE of links from Rx1 to Tx1, followed by links to Tx2, then Tx3, and then Tx4 are 19.7%, 3.25%, 12.3%, and 0.63%, respectively. From these results, it is obvious that channels with strong signal strength and DC channel gain above -35 dB have lower MSE, i.e. less than 5%.

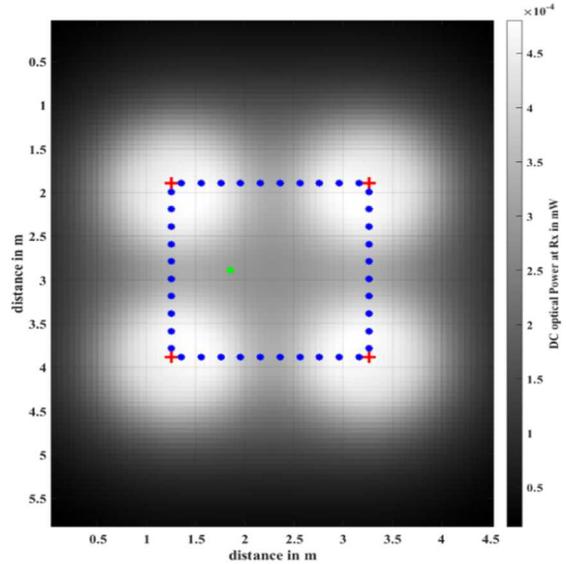

Fig. 4. Amplitude optical power distribution. All transmitters are marked in red color. Positions of Rx1 are marked in blue color and Rx2 position is marked in green.

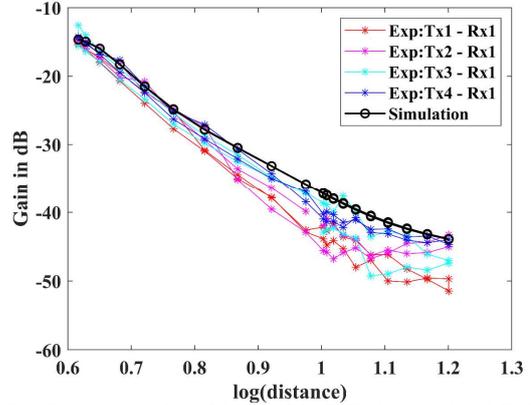

Fig. 5. Channel gain at 5 MHz of the LiFi channels in a MIMO mobility scenario.

Fig. 3 (b) shows the channel responses for Rx2, placed in the middle between Tx1 and Tx3. It is clear that Rx2 has strong LOS signals from Tx1 and Tx3 and weak signals from Tx2 and Tx4. Here, in the simulated channel responses, the links between Rx2 to Tx1 and Rx2 to Tx3 will have the same response. Similarly, links between Rx2 to Tx2 and Rx2 to Tx4 are similar. As explained before, due to mismatch in the optical frontends and other connectors, there will always be minor deviations in the measurement results, which are not identical to those in the simulations. Here MSE of links from Rx2 to Tx1, followed by links to Tx2, then Tx3, and then Tx4 are 1.26%, 24%, 0.96%, and 36%, respectively. We observed that MSE of the links with high channel gain is lower.

*C. MIMO Mobility Scenario*

In this scenario, we consider that Rx1 is moving around the room while another receiver Rx2 is kept at a fixed position. Fig. 4 shows the calculated heat map of the expected DC optical power distribution in the room where Rx is kept at a height of 1m. Positions of all Tx are marked in red color and positions of Rx1 and Rx2 are marked in blue and green colors, respectively. We have done measurements by moving Rx1 to 40 different positions as shown in Fig. 4.

To compare the experimental data with the simulations, we calculate the channel gain at a lower frequency of 5 MHz corresponding to each position. Fig. 5 shows the variation of the channel gain concerning the logarithmic distance of the separation between Tx and Rx. Channel gain is plotted for each Tx-Rx link for all 40 positions. When the receiver is moving far away from good illumination coverage of one of the transmitter, then the corresponding channel gains are reduced. In the experiment, we observed that channel gain variation is between -15 dB to -48 dB for distance variation from 1.85m to 3.33m. Due to mismatches in the optical frontends, there will be negligible differences in the channel gains at lower distances. When Rx1 is far from the transmitters, then corresponding channel gains will be lower and there is random variation due to noisy data.

In the simulations, since all transmitters are placed in a 2 m x 2 m grid, channel gain variations for all Tx links with respect to Rx1 will be the same. As shown in Fig. 5, the channel gain variations for the simulated channel gain at 5 MHz is from -15 dB to -44 dB. When compared with experimental data, the deviation of simulated channel gain with respect to experimental data is less at lower distances. The difference in simulated channel gain, when compared to experimental data is significant for higher distance where channel gain is lower than -35 dB. This study shows that approximate results for channel gain variations can be estimated for a mobile device with low error. From the numerically calculated channel, we can calculate MIMO channel characteristics, such as statistics of singular values and channel throughput with small error and compare them with experimental data.

## V. DISCUSSION

This study proposes an efficient channel modelling method for LiFi channels, by combining the frequency-domain technique in [6] with the integrating sphere approach in [7]. Our results indicate a good agreement between simulation results and experimental results. As principal limitations, we have considered a simplified geometrical model of the room to verify this technique first in a scenario with minor complexity and characterize the new channel modelling method in a mobile scenario in this way. Since we are calculating the channel transfer functions, gain variations can be calculated much faster than using ray tracing in the time domain.

The implementation complexity of the new method depends mostly on the performance of the selected computer. The mobile scenario was calculated in a few minutes on a standard laptop. Since most calculations are based on matrix multiplications, instead of tracing each ray individually, it is possible to speed up the calculations further by utilizing high performance GPUs. This approach can be suitable to estimate models especially for crowded LiFi systems with large numbers of mobile users. It may also be helpful for estimating how many access points are visible, how to combine signals for optimized communication and characterize positioning in large LiFi deployments.

Effective communication in most of the LiFi channels is either due to LOS or first order reflections. While the user is moving in the room, the propagation channels vary. Using our simulation methodology, we can easily estimate approximate results for these channel variations. We demonstrate that simulated channel responses with gain above -35 dB have 1 to 2 dB differences when compared with experimental results. As the gain goes below -35 dB, there is significant noise in the experimental data and observe the significant difference of around 2 to 4 dB when compared with simulation.

Note that, in this simulation model, as a major limitation, accuracy of the method depends mainly on the size and number of surface elements. As the size gets smaller, there will be more surface elements and thus computation time gets increases. The relation between resolution of surface element $\Delta x$ and time resolution $\Delta t$ of the impulse response can be expressed as $\Delta x = c \cdot \Delta t$, where $c$ is the speed of the light. With better time resolution, smaller i.e. more surface elements are required, and hence the matrices in Equation (2) get larger and simulation time will increase.

Moreover, this simulation model can be used for other indoor scenarios including rooms filled with more obstacles. In that case, using other algorithms, e.g. [20], possible blockages in each link should be captured exactly and included in the LOS as well as in frequency domain model. In the case of diffuse reflections, the first two reflections usually result in distinct peaks in the impulse response while all the later diffuse reflections fold into one exponential decay (see Fig.3 in [7]). Diffuse reflections from additional objects in the room will be included precisely by using the frequency-domain technique and their overall impact on the path loss and the average time-of-flight will be captured by the sphere model.

## VI. CONCLUSION

In this paper, a simplified numerical channel modelling method for the indoor LiFi communication channel has been presented. The LiFi channel simulations for LOS as well as NLOS diffuse paths are computed in the frequency domain rather than in time domain. For NLOS diffuse channels, we have considered the frequency domain channel modelling technique along with the integrating sphere model.

At first, to validate our simulation method, we compared the simulation results with the previously published SISO measurement results. These results show that there exists a good matching when compared with experimental data with minimum error. To validate our simulation method in a realistic LiFi scenario, we conducted the 4x2 distributed MIMO measurement in an empty room. Then we compared the measurement results with simulations results and show that there is a good agreement between the results.

In the same MIMO configuration, we considered one Rx at 40 different positions in the room. Measurement and simulation results indicate that channels with a gain of more than -35dB are in good agreement with respect to experimental results and channels with gain less than -35 dB have a difference of 2-4 dB. The major advantage of our new modelling approach is the reduced computation time, compared to the ray tracing, which allows the efficient modelling of large LiFi scenarios with many mobile devices, suitable for future IoT applications.

## ACKNOWLEDGMENT

This research is funded by the VisIoN project, a European Union's H-2020 MSCA ITN program under the grant agreement no 764461. It is also based upon work from COST Action CA19111 NEWFOCUS, supported by COST (European Cooperation in Science and Technology).